\documentclass[showkeys,showpacs,prd]{revtex4}
\usepackage{amsmath}
\usepackage{amssymb}
\usepackage{graphicx}
\usepackage{color}

\begin{document}

\title{
Modified gravity from the nonperturbative quantization of a metric
}

\author{
Vladimir Dzhunushaliev,$^{1,2,3,4,5}$
\footnote{Email: v.dzhunushaliev@gmail.com}
Vladimir Folomeev,$^{2,3,4}$
\footnote{Email: vfolomeev@mail.ru}
Burkhard Kleihaus,$^{4}$
\footnote{Email: b.kleihaus@uni-oldenburg.de }
Jutta Kunz$^4$
\footnote{Email:  jutta.kunz@uni-oldenburg.de}
}
\affiliation{$^1$
Dept. Theor. and Nucl. Phys., Al-Farabi Kazakh National University, Almaty, 050040, Kazakhstan \\
$^2$ IETP, Al-Farabi Kazakh National University, Almaty, 050040, Kazakhstan \\
$^3$Institute of Physicotechnical Problems and Material Science of the NAS
of the Kyrgyz Republic, 265~a, Chui Street, Bishkek, 720071,  Kyrgyz Republic \\
$^4$Institut f\"ur Physik, Universit\"at Oldenburg, Postfach 2503
D-26111 Oldenburg, Germany \\
$^5$Institute for Basic Research,
Eurasian National University,
Astana, 010008, Kazakhstan
}

\begin{abstract}
Based on certain assumptions for the expectation value of a product of the quantum fluctuating metric at two points, the gravitational and scalar field Lagrangians are  evaluated. Assuming a vanishing expectation value of the first order terms of the metric, the calculations are performed with an accuracy of second order. It is shown that such quantum corrections give rise to modified gravity.
\end{abstract}


\maketitle

\section{Introduction}

The problem of quantizing gravity has been debated frequently during the past decades. In doing so, it was established that the quantization of Einstein's general relativity inevitably results in fundamental problems such as the perturbative nonrenormalizability~\cite{Birrell}. This represents a motivation to introduce other, more radical approaches to obtain a quantum theory of gravity, including the consideration of higher-order theories of gravity, string theory, and loop quantum  gravity~\cite{Papantonopoulos:2013}.

One possibility is to reject the perturbative quantization, as is done, for example, within the framework of loop-space nonperturbative quantum gravity \cite{Rovelli:1991zi}. On the other hand, in order to quantize gravity one can try to adopt the nonperturbative quantization technique employed by Heisenberg when considering a nonlinear spinor field theory~\cite{heis}. The central idea of this approach is that the description of the quantum system is achieved by using an infinite set of equations for all Green's functions.
Based on physically motivated arguments, one can introduce a cutoff procedure to obtain a finite number of equations. Such a procedure is close to the one
used in turbulence modeling~\citep{Wilcox}.

Working within this approach, we here consider the situation where a quantum metric can be decomposed into a sum of an averaged (classical) metric
$g_{\mu \nu}$ and a fluctuating (quantum) part $\delta g_{\mu \nu}$. Also, we assume that the Green's function of a product of the fluctuating part of the metric can be approximated in a certain way. Under these assumptions we calculate the action for the gravitational field and for matter (in the form of a scalar field) with an accuracy of $\delta g^2$. We show that the resulting action represents modified $F(R)$-type theories of gravity, which are now widely used in modeling the present accelerated expansion of the Universe~\cite{Nojiri:2010wj}.

For the quantum gravitating physical system considered here, with the decomposition of the metric into classical and quantum parts, the expectation value of the quantum part can be taken to be zero or nonzero, based on certain physical motivations. In the present paper we take it to be zero. This is in contrast to our previous work~\cite{Dzhunushaliev:2013nea}, where the decomposition into two parts (classical and quantum) was also performed,
but the expectation value of the quantum part of the metric was taken to be nonzero. Therefore, we here take the second variation of the metric into
account, as discussed below.

\section{Nonperturbative quantization technique}

According to Heisenberg's nonperturbative quantization technique the classical fields appearing in the corresponding field equations are replaced by operators of these fields. For general relativity, one then has the operator Einstein equations
\begin{equation}
\hat G_{\mu \nu}\equiv	\hat R_{\mu \nu} - \frac{1}{2} \hat g_{\mu \nu} \hat R =
	\frac{8\pi G}{c^4} \hat T_{\mu \nu} ,
\label{1-10}
\end{equation}
where all geometric operators
$\hat R_{\mu \nu}$,
$\hat R^\rho_{\phantom{\rho} \sigma \mu \nu}$,
$\hat \Gamma^\rho_{\phantom{\rho} \mu \nu}$
are defined in the same way as in the classical case, and differ only in the replacement of the classical quantities by the corresponding operators (for details, see Ref.~\cite{Dzhunushaliev:2013nea}).

There are no known mathematical tools for solving the operator equation \eqref{1-10}. The only possibility to work with such an operator equation is to average the equation \eqref{1-10} over all possible products of the metric operators $\hat g(x_1) \cdots \hat g(x_n)$, and thus obtain an infinite set of equations for all Green's functions:
\begin{eqnarray}
	\left\langle Q \left| \hat g (x_1)
	\cdot \hat G_{\mu \nu}
	\right| Q \right\rangle &=& \frac{8\pi G}{c^4}
  \left\langle Q \left| \hat g (x_1)
	 \cdot \hat T_{\mu \nu}
	\right| Q \right\rangle ,
\label{1-50}\\
	\left\langle Q \left| \hat g(x_1) \hat g(x_2)
	\cdot \hat G_{\mu \nu}
	\right| Q \right\rangle &=& \frac{8\pi G}{c^4}
  \left\langle Q \left| \hat g (x_1) \hat g (x_2)
	 \cdot \hat T_{\mu \nu}
	\right| Q \right\rangle ,
\label{1-60}\\
	\cdots &=& \cdots 	,
\label{1-150}\\
	\left\langle Q \left|
	\text{ the product of $g$ at different points $(x_1,
	\cdots , x_n)$} \cdot \hat G_{\mu \nu}
	\right| Q \right\rangle &=&
\nonumber \\
  \frac{8\pi G}{c^4}
  \left\langle Q \left|
    \text{ the product of $g$ at different points $(x_1,
	  \cdots , x_n)$}
	   \cdot \hat T_{\mu \nu}
	\right| Q \right\rangle,
&&
\label{1-70}
\end{eqnarray}
where $\left. \left|Q \right. \right\rangle$ is a quantum state (for details
see Refs.~\cite{Dzhunushaliev:2013nea,Dzhunushaliev:2012np}). The exact definitions of a non-perturbative vacuum state $\left. \left. \right| 0 \right\rangle$ and a quantum state $\left. \left. \right| Q \right\rangle$ are given in Appendix \ref{NPvacuum}.

Evidently Eqs.~\eqref{1-50}-\eqref{1-70} cannot be solved analytically.
Different possibilities to solve them approximately were discussed in
Refs.~\cite{Dzhunushaliev:2013nea,Dzhunushaliev:2012np,Dzhunushaliev:2012vb}.
Note, that similar mathematical problems appear in turbulence modeling, where an infinite set of equations for all cumulants also arises~\cite{Wilcox}.

Here we use the following strategy for approximate solving
Eqs.~\eqref{1-50}-\eqref{1-70}: we decompose the metric operator
$\hat g_{\mu \nu}$ into classical, $g_{\mu \nu}$,
and quantum, $\widehat{\delta g}_{\mu \nu}$, parts and evaluate the expectation value of the Lagrangian with an accuracy
$\left\langle \left( \widehat{\delta g} \right)^2 \right\rangle$.
This strategy is similar to the one employed in connection with quantum torsion in Ref.~\cite{Dzhunushaliev:2012vb}.

\section{Assumptions of the quantum averaging}

In accordance with the quantization procedure, a metric in quantum gravity is an operator $\hat g_{\mu\nu}$. Here we consider a system for which the following decomposition is approximately valid
\begin{equation}
	\hat g_{\mu\nu} \approx g_{\mu \nu} + \widehat{\delta g}_{\mu \nu} +
	\widehat{\delta^2 g}_{\mu \nu},
\label{2-10}
\end{equation}
where $g_{\mu \nu}$ is the classical part of the metric;
$
\left\langle Q \left|
	\widehat{\delta g}_{\mu \nu}
\right | Q \right\rangle = 0
$;
$
\left\langle Q \left|
	\widehat{\delta^2 g}_{\mu \nu}
\right | Q \right\rangle \neq 0
$;
$\left. \left. \right |Q \right\rangle$ is some quantum state;
$\widehat{\delta g}_{\mu \nu}$ and $\widehat{\delta^2 g}_{\mu \nu}$
are the first- and second-order deviations of the operator $\hat g_{\mu\nu}$.
The expression \eqref{2-10} is a quantum variant of the decomposition
of the classical metric (see Ref.~\cite{Besse:2007}, p.~129),
where $\widehat{\delta g}_{\mu \nu}$ is the first order term
and $\widehat{\delta^2 g}_{\mu \nu}$ is the second order term.
In our quantum case it means that
$ \left\langle \left( \widehat{\delta g} \right)^2 \right\rangle
\approx \left\langle \widehat{\delta^2 g} \right\rangle $.

In Ref.~\cite{Dzhunushaliev:2013nea} we have considered
a physical quantum system with the decomposition
$ \hat g_{\mu \nu} =g_{\mu \nu} + \widehat{\delta g}_{\mu \nu} $,
where $g_{\mu \nu}$ is the classical part and
$\widehat{\delta g}_{\mu \nu}$ is the quantum part of the metric
with the nonzero vacuum expectation value
$ \left\langle \widehat{\delta g}_{\mu \nu} \right\rangle $.
Thus the difference between the physical system
of Ref.~\cite{Dzhunushaliev:2013nea} and the one of the present paper
is that in the first case the final effective Lagrangian
is calculated with accuracy
$ \left\langle \widehat{\delta g}_{\mu \nu} \right\rangle \neq 0 $,
and in the second case the final averaged Lagrangian is calculated
with  accuracy
$
\left\langle
  \left( \widehat{\delta g}_{\mu \nu} \right)^2
\right\rangle
$ and
$
\left\langle
  \widehat{\delta^2 g}_{\mu \nu}
\right\rangle
$.

For our approximate nonperturbative calculations,
we insert the decomposition \eqref{2-10} into the Einstein-Hilbert action
and evaluate it with an accuracy
$
\left\langle \left( \widehat{\delta g} \right)^2 \right\rangle
\approx \left\langle \widehat{\delta^2 g} \right\rangle
$.
To do this, we have to make some assumptions on the 2-point Green's function.
Namely, we suppose that it can be decomposed as the product
\begin{equation}
	G_{2; \mu \nu, \rho \sigma} \left( x_1, x_2 \right) =
	\left\langle Q \left|
		\widehat{\delta g_{\mu \nu}}(x_1) \cdot \widehat{\delta g_{\rho \sigma}}(x_2)
	\right | Q \right\rangle.
\label{2-20}
\end{equation}
It is seen from this expression
that $G_2$ should be symmetric under the permutations
$\mu \leftrightarrow \nu$, $\rho \leftrightarrow \sigma$,
and $\mu, \nu \leftrightarrow \rho, \sigma$.
In the subsequent calculations we will use the following assumptions:
\begin{itemize}
\item The Green's function $G_2(x_1, x_2)$ can be approximately decomposed
as the product of some tensors at the points $x_1$ and $x_2$:
\begin{equation}
	G_{2; \mu \nu, \rho \sigma} \left( x_1, x_2 \right) \approx
	P_{\mu \nu}(x_1) P_{\rho \sigma}(x_2).
\label{2-30}
\end{equation}
\item Taking into account the symmetry properties noted above,
one can see that there exist the following possibilities for
choosing the tensor $P_{\mu \nu}$:
	\begin{itemize}
	\item $P_{\mu \nu}$  is proportional to the metric tensor:
	\begin{equation}
		P_{\mu \nu} \propto g_{\mu \nu}.
	\label{2-40}
	\end{equation}	
	\item $P_{\mu \nu}$  is proportional to the Ricci tensor:
	\begin{equation}
		P_{\mu \nu} \propto \frac{R_{\mu \nu}}{R}.
	\label{2-50}
	\end{equation}		
	\end{itemize}
	\item The proportionality coefficient in the expressions
\eqref{2-40} and \eqref{2-50} should be some invariant.
Consequently, it has to have the form $F(R, R_{\mu \nu} R^{\mu \nu}, \cdots)$.
The coefficient $F$ should be very small
as $g_{\mu \nu} \rightarrow \eta_{\mu \nu}$,
where $\eta_{\mu \nu}$ is the Minkowski metric.	
	\item The expectation value
$\left\langle \widehat{\delta^2 g_{\mu \nu}}  \right\rangle$
is given by some tensor of rank two,
	\begin{equation}
		\left\langle \widehat{\delta^2 g_{\mu \nu}} \right\rangle
		= K_{\mu \nu}.
	\label{2-55}
	\end{equation}
	\item Analogously to \eqref{2-40} and \eqref{2-50}, we assume that
	\begin{itemize}
    	\item $K_{\mu \nu}$  can be proportional to the metric tensor:
    	\begin{equation}
    		K_{\mu \nu} \propto g_{\mu \nu}.
    	\label{2-56}
    	\end{equation}	
    	\item $K_{\mu \nu}$  can be proportional to the Ricci tensor:
    	\begin{equation}
    		K_{\mu \nu} \propto \frac{R_{\mu \nu}}{R}.
    	\label{2-57}
	\end{equation}		
	\end{itemize}
  \item $P_{\mu \nu}$ and $K_{\mu \nu}$ can be a linear combination
of the metric and Ricci tensors.
For example, they could be the Einstein or Schouten tensors.		
  \item For each quantum state $\left| Q \right\rangle$ there exists
only a single set of functions $F$ and $K_{\mu \nu}$.
\end{itemize}
Thus, we assume that the quantum correlation between fluctuations
of the metric at two points can be approximately described as
\begin{equation}
	G_{2; \mu \nu, \rho \sigma} \left( x_1, x_2 \right) \approx
	\left[
	P_{\mu \nu}  F(R, R_{\mu \nu} R^{\mu \nu}, \cdots)
	\right]_{x_1} \cdot
	\left[
	P_{\rho \sigma}   F(R, R_{\mu \nu} R^{\mu \nu}, \cdots)
	\right] _{x_2}.
\label{2-60}
\end{equation}
The simplest choice for $F$ is
\begin{equation}
	 F =  F(R),
\label{2-70}
\end{equation}
which corresponds to $F(R)$-gravities.

\section{Evaluation of the averaged action}

We start from the classical Einstein-Hilbert Lagrangian
\begin{equation}
	\mathcal L = - \frac{c^2}{2 \varkappa} \sqrt{-g} R,
\label{3-10}
\end{equation}
where $\varkappa=8\pi G/c^2$. We then expand
$\mathcal L(g + \delta g+\delta^2 g)$ into a Taylor series
and subsequently replace the classical quantities $\delta g, \delta^2 g$
by quantum ones
$\widehat{\delta g}, \widehat{\delta^2 g}$
\begin{equation}
	\hat{\mathcal L}(g + \widehat{\delta g}+\widehat{\delta^2 g}) \approx \mathcal L(g) +
	\frac{\delta \mathcal L}{\delta g^{\mu \nu}} \widehat{\delta g^{\mu \nu}} +
	\frac{\delta^2 \mathcal L}{\delta g^{\mu \nu} \delta g^{\rho \sigma}}
	\widehat{\delta g^{\mu \nu}} \widehat{\delta g^{\rho \sigma}} +
	\frac{\delta^2 \mathcal L}{\delta^2 g^{\mu \nu}}
	\widehat{\delta^2 g^{\mu \nu}} .
\label{3-20}
\end{equation}
Our next step is to average this Lagrangian over quantum fluctuations
of the metric $\delta g_{\mu \nu}$. Since in the present paper we assume
that $\left\langle \widehat{\delta g^{\mu \nu}} \right\rangle = 0$
(see the Introduction and cf. Ref.~\cite{Dzhunushaliev:2013nea}
where it was taken to be nonzero),
we have
\begin{equation}
	\left\langle 	
		\delta^2 \hat{\mathcal L}(g )
	\right\rangle
	= 	\frac{\delta^2 \mathcal L(g)}{\delta g^{\mu \nu} \delta g^{\rho \sigma}}
	\left\langle 		
		\widehat{\delta g^{\mu \nu}} \widehat{\delta g^{\rho \sigma}}
	\right\rangle +
	\frac{\delta^2 \mathcal L}{\delta^2 g^{\mu \nu}}
	\left\langle 		
		\widehat{\delta^2 g^{\mu \nu}}
	\right\rangle.
\label{3-30}
\end{equation}
Taking into account the first variation of the Einstein-Hilbert Lagrangian
\begin{equation}
	\delta \mathcal L(g ) = - \frac{c^2}{2 \varkappa} \sqrt{-g} \left(
	 R_{\mu \nu} - \frac{1}{2} g_{\mu \nu} R
	\right) \delta g^{\mu \nu},
\label{3-40}
\end{equation}
we can calculate the second variation as follows:
%of the Einstein - Hilbert Lagrangian
\begin{equation}
	\delta^2 \mathcal L(g ) = - \frac{c^2}{2 \varkappa} \left[
	\left( \delta{\sqrt{-g}} \right)
	\left(
	 R_{\mu \nu} - \frac{1}{2} g_{\mu \nu} R
	\right) 	+
	\sqrt{-g} \delta \left(
	 R_{\mu \nu} - \frac{1}{2} g_{\mu \nu} R
	\right) \right]  \delta g^{\mu \nu} +
	\sqrt{-g}\, G_{\mu \nu} \delta^2 g^{\mu \nu},
\label{3-50}
\end{equation}
where $G_{\mu \nu} = R_{\mu \nu} - \frac{1}{2} g_{\mu \nu}R$
is the Einstein tensor. Equation \eqref{3-50} can be rewritten as
\begin{equation}
%\begin{split}
	\delta^2 \mathcal L(g ) = - \frac{c^2}{2 \varkappa} \sqrt{-g} \left\lbrace
	\left[
	- \frac{1}{2} \left(
	 R_{\mu \nu} - \frac{1}{2} g_{\mu \nu} R
	\right) g_{\alpha \beta} 	\delta g^{\alpha \beta}  +
	 \delta R_{\mu \nu}  -
	 \frac{R}{2} \delta g_{\mu \nu} -
	 \frac{1}{2} g_{\mu \nu} \delta R
	\right] \delta g^{\mu \nu} + G_{\mu \nu} \delta^2 g^{\mu \nu}
	\right\rbrace.
\label{3-60}
%\end{split}
\end{equation}
Now we can calculate an expectation value of the Lagrangian \eqref{3-20}
by replacing all classical quantities $\delta g^{\mu \nu}$
and $\delta^2 g^{\mu \nu}$ in Eq.~\eqref{3-60} by the quantum ones,
$\widehat{\delta g}^{\mu \nu}$ and $\widehat{\delta^2 g}^{\mu \nu}$
(for details see Appendix~\ref{app}).
For simplicity, let us here consider the Ansatz \eqref{2-40},
for which we obtain
\begin{equation}
	\left\langle \hat{\mathcal L}(g + \widehat{\delta g}+\widehat{\delta^2 g}) \right\rangle \approx
	- \frac{c^2}{2 \varkappa} \sqrt{-g} \left[ R -
	2 R F(R, \cdots) + 3 F(R, \cdots) \nabla^\mu \nabla_\mu F(R, \cdots) +
  G_{\mu \nu} K^{\mu \nu}
	\right].
\label{3-100}
\end{equation}
Thus, we see that we have derived a modified gravity theory.
For the simplest choice $F(R, \cdots) = F(R)$ and $K_{\mu \nu} = 0$
we have $F(R)$-gravity theory.

Let us now perform similar calculations for the matter Lagrangian,
using the decomposition given by Eqs.~\eqref{2-10} and \eqref{2-20}.
For simplicity,  consider the scalar field $\phi$ with the Lagrange density
\begin{equation}
	\mathcal L_m = \sqrt{-g}L_m=\sqrt{-g}
	\left[ 	
		\frac{1}{2}\nabla^\mu \phi \nabla_\mu \phi - 	V(\phi)
	\right] .
\label{3-110}
\end{equation}
Its first variation is
\begin{equation}
	 \delta \mathcal L_m = \frac{\sqrt{-g}}{2}
	\left[		
		\nabla_\mu \phi \nabla_\nu \phi -
		g_{\mu \nu}  L_m
	\right] \delta g^{\mu \nu} =
	\frac{\sqrt{-g}}{2} T_{\mu \nu} \delta g^{\mu \nu},
\label{3-120}
\end{equation}
where $T_{\mu \nu}$ is the energy-momentum tensor.
Then the second variation yields
\begin{equation}
	 \delta^2 \mathcal L_m = \frac{\sqrt{-g}}{2} \left\lbrace
	\left[
	\left( 		
		-g_{\mu \alpha} g_{\nu \beta} + \frac{1}{2}g_{\mu \nu} g_{\alpha \beta}
	\right)  L_m -
	g_{\alpha \beta} \nabla_\mu \phi \nabla_\nu \phi
	\right] \delta g^{\mu \nu} \delta g^{\alpha \beta} +
	T_{\mu \nu} \delta^2 g^{\mu \nu}
	\right\rbrace.
\label{3-130}
\end{equation}
Using Eqs.~\eqref{2-20}-\eqref{2-40} and \eqref{2-55}
and replacing again all classical quantities
$\delta g^{\mu \nu}$ and $\delta^2 g^{\mu \nu}$ by the quantum ones,
we find the expectation value
$
\left\langle \mathcal L_m + \widehat{\delta^2 \mathcal L_m} \right\rangle
$ in the form
\begin{equation}
	\left\langle \mathcal L_m + \widehat{\delta^2 \mathcal L_m} \right\rangle =
	\sqrt{-g}\left\{
		\frac{1}{2}
		 \nabla^\mu \phi \nabla_\mu \phi -
		\left[
			1 + 2 F\left( R, \cdots \right)
		\right] V(\phi) +
		K^{\mu \nu} T_{\mu \nu}
	\right\} .
\label{3-140}
\end{equation}
Thus, we see that
a nonminimal coupling between the scalar field and gravity appears.
Notice also, that in the case of
$K_{\mu \nu} \neq 0$ one can obtain a gravitational theory
in which the derivative of the scalar field $\phi$,
appearing in the term $T_{\mu \nu}$,
is nonminimally coupled to curvature
(for cosmological models with such a coupling,
see, e.g., Refs.~\cite{Ame,CapLamSch,Sushkov:2009hk}).

\section{Conclusion}

In the present paper we have considered the case of
a quantum gravitating system when the metric can be decomposed into
classical and quantum parts. For such a system,
we have calculated the gravitational and matter Lagrangians with
an accuracy up to the second variation of the metric. In doing so,
we have decomposed the operator of the metric into a sum
of its expectation value ($c$-number)
and deviations ($q$-numbers) from this expectation value.

Based on certain assumptions on the dispersion of quantum fluctuations
of the metric, we have shown that:
\begin{itemize}
\item Einstein gravity is modified in the spirit of $F(R)$-gravity theories.
\item Matter is nonminimally coupled to gravity.
\end{itemize}

In obtaining these results,
we have assumed that the expectation value of the product
$\left\langle \widehat{\delta g_{\mu \nu}}(x_1)
\widehat{\delta g_{\mu \nu}} (x_2) \right\rangle $ at two points $x_1, x_2$
can be decomposed into the product of two factors of
some tensor at these two points,
see Eqs.~\eqref{2-20} and \eqref{2-30}.

The proposed model, which takes into account
the quantum fluctuations of the metric,
provides us with the following scheme
for the explanation of the present acceleration of the Universe:
the quantum metric $\rightarrow F(R)$-gravity $\rightarrow$
the accelerated Universe.
The model can explain qualitatively the smallness
of the effective $\Lambda$-term,
which comes from small quantum fluctuations of the metric.

Note that the proposed procedure of quantizing the metric
may be considered as being related to quantum gravity
like the phenomenological Ginzburg-Landau model of superconductivity
is related to the microscopical Bardeen-Cooper-Schrieffer theory:
the proportionality coefficient $F(R, \cdots)$ can be calculated only
from the true quantum gravity.

\section*{Acknowledgements}
We are grateful to the Volkswagen Stiftung for the support of this research.
VD and VF acknowledge support from a grant No.~0263/PCF-14 in fundamental research in natural sciences by the Ministry of Education and Science of Kazakhstan. They also would like to thank the Carl von Ossietzky University of Oldenburg for hospitality while this work was carried out. BK and JK gratefully acknowledge support by the DFG Research Training Group 1620 ``Models of Gravity'', and by FP7, Marie Curie Actions, People, International Research Staff Exchange Scheme (IRSES-606096).

\appendix

\section{Variation of geometrical quantities}
\label{app}

Following Ref.~\cite{Lichnerowicz},
we here give all formulae concerning the variation of
$R, R_{\mu \nu}$, and $R_{\mu \nu \alpha \beta}$.
The variations of the inverse metric and
the Christoffel symbols $\Gamma^\mu_{\nu \rho}$ are
\begin{eqnarray}
	\delta g^{\alpha \beta} &=&
	- g^{\alpha \rho} g^{\beta \sigma} \delta g_{\rho \sigma}
	= - h^{\alpha \beta},
\label{a1-5}\\
	\delta \Gamma_{\gamma \alpha \beta} &=&
	\frac{1}{2} \left(
		\nabla_\alpha h_{\beta \gamma} +
		\nabla_\beta h_{\gamma \alpha}-
		\nabla_\gamma h_{\alpha \beta}
	\right) + h_{\gamma \rho} \Gamma^{\rho}_{\alpha \beta},
\label{a1-10}\\
	\delta \Gamma^{\mu}_{\nu \rho} &=&
	\frac{1}{2} \left(
		\nabla_\nu \delta g^\mu_\rho +
		\nabla_\nu \delta g^\mu_\rho -
		\nabla^\mu \delta g^\nu_\rho
	\right),
\label{a1-20}
\end{eqnarray}
where $g_{\mu \nu}$ and $g^{\mu \nu}$ are used to lower and raise indices,
and for brevity, we have introduced $h_{\mu \nu} = \delta g_{\mu \nu}$.
Then the variation of the Riemann and Ricci tensors
and the curvature scalar are
\begin{eqnarray}
	\delta R^\alpha_{\phantom{\alpha} \beta \gamma \delta} &=&
	\nabla_\gamma \delta \Gamma^\alpha_{\beta \delta} -
	\nabla_\delta \delta \Gamma^\alpha_{\beta \gamma},
\label{a1-25}\\
	\delta R_{\alpha \beta \gamma \delta} &=& \frac{1}{2}
	\left[
		h_{\alpha\rho} R^\rho_{\phantom{\rho} \beta\gamma \delta} +
		h_{\beta\rho} R^{\phantom{\alpha} \rho}_{\alpha \phantom{\rho} \gamma \delta} -
		\left(
			\nabla_\delta \nabla_\beta h_{\gamma \alpha} +
			\nabla_\gamma \nabla_\alpha h_{\delta \beta} -
			\nabla_\gamma \nabla_\beta h_{\delta \alpha} -
			\nabla_\delta \nabla_\alpha h_{\gamma \beta}
		\right)
	 \right] ,
\label{a1-30}\\
	\delta R_{\alpha \beta} &=&
	\frac{1}{2} \left[
	 - \nabla^\rho \nabla_\rho h_{\alpha \beta} +
	 R^{\phantom{\alpha} \sigma}_{\alpha} h_{\sigma \beta} +
	 R^{\phantom{\beta} \sigma}_{\beta} h_{\sigma \alpha} -
	 2 R_{\alpha \rho \beta \sigma} h^{\rho\sigma} +
	 \nabla_\alpha \nabla_\rho h_\beta^{\phantom{\beta} \rho} +
	 \nabla_\beta \nabla_\rho h_\alpha^{\phantom{\alpha} \rho} -
	 \nabla_\alpha \nabla_\beta h
	\right] ,
\label{a1-40}\\
	\delta R &=& h^{\alpha \beta} R_{\alpha \beta} + 	
	g^{\alpha \beta} \delta R_{\alpha \beta} =
	h^{\alpha \beta} R_{\alpha \beta} - \nabla^\alpha \nabla_\alpha h +
	\nabla^\alpha \nabla^\beta h_{\alpha \beta},
\label{a1-50}
\end{eqnarray}
where $h= g^{\alpha \beta} h_{\alpha \beta}$
and the covariant derivative $\nabla_\mu$ is taken
with respect to the metric $g$.

Let us now calculate the expectation values for the case \eqref{2-40}:
\begin{eqnarray}
	- \frac{1}{2} \left(
		R_{\mu \nu} - \frac{1}{2} g_{\mu \nu} R
	\right) g_{\alpha \beta} \left\langle
		\widehat{\delta g^{\mu \nu}} \widehat{\delta g^{\alpha \beta}}
	\right\rangle &=& 2 R  F^2(R, \cdots) ,
\label{a1-70}\\
	\frac{1}{2} \left\langle
		\widehat{\delta R_{\mu \nu}} \widehat{\delta g^{\mu \nu}}
	\right\rangle &=& - 3 F \nabla^\mu \nabla_\mu  F,
\label{a1-90}\\
	- \frac{1}{2} R \left\langle
	\widehat{\delta g_{\mu \nu}} \widehat{\delta g^{\mu \nu}}
	\right\rangle &=& -2 R  F^2(R, \cdots) ,
\label{a1-100}\\
	- \frac{1}{2} g_{\mu \nu} \left\langle
	\widehat{\delta R} \widehat{\delta g^{\mu \nu}}
	\right\rangle &=& - 2 R  F^2(R, \cdots) +
	6  F \nabla^\mu \nabla_\mu  F,
\label{a1-110}
\end{eqnarray}
where we have used the fact
\begin{equation}
	\left\langle
		\left( \nabla_\mu \widehat{\delta g_{\alpha \beta}} \right)
		\widehat{\delta g_{\rho \sigma}}
		\right\rangle = \lim \limits_{x_2 \rightarrow x_1}
		\left( \nabla_\mu\right)_{x_1}
		\left\langle
			\widehat{\delta g_{\alpha \beta}} (x_1)
		\widehat{\delta g_{\rho \sigma}} (x_2)
		\right\rangle.
\label{a1-120}
\end{equation}
Summing \eqref{a1-70}-\eqref{a1-110} yields
\begin{equation}
	\left\langle
		\widehat{\delta^2 \mathcal L}
	\right\rangle = \frac{c^2}{\varkappa} \sqrt{-g} R  F(R, \cdots).
\label{a1-130}
\end{equation}
For the case \eqref{2-50}, we obtain the following expectation values:
\begin{eqnarray}
	- \frac{1}{2} \left(
		R_{\mu \nu} - \frac{1}{2} g_{\mu \nu} R
	\right) g_{\alpha \beta} \left\langle
		\widehat{\delta g^{\mu \nu}} \widehat{\delta g^{\alpha \beta}}
	\right\rangle &=& -\frac{1}{2} \left(
		R_{\alpha \beta }R^{\alpha \beta} -
		\frac{1}{2} R^2
	\right) \frac{F^2}{R},
\label{a1-140}\\
	\frac{1}{2} \left\langle
		\widehat{\delta R_{\mu \nu}} \widehat{\delta g^{\mu \nu}}
	\right\rangle &=& \frac{1}{4} \left[
		-  F \frac{R^{\alpha \beta}}{R} \left(
			\nabla^\rho \nabla_\rho \frac{F R_{\alpha \beta}}{R}
		\right) +
		2 R_\alpha^\sigma R_{\sigma \beta} R^{\alpha \beta}
		\frac{F^2}{R} -
	\right.
\nonumber \\
	&&
	\left.
		2 R_{\alpha \rho \beta \sigma} R^{\rho \sigma} R^{\alpha \beta} \frac{F^2}{R^2} +
		2 F \frac{R^{\alpha \beta}}{R} \nabla_\alpha \nabla_\rho \left(
			F \frac{R_\beta^\rho}{R}
		\right)  -
	\right.
\nonumber \\
		&&
	\left.
		 F \frac{R^{\alpha \beta}}{R} \nabla_\alpha \nabla_\beta  F
	\right] ,
\label{a1-150}\\
	- \frac{1}{2} R \left\langle
	\widehat{\delta g_{\mu \nu}} \widehat{\delta g^{\mu \nu}}
	\right\rangle &=& - \frac{F^2}{2} \frac{R_{\alpha \beta }R^{\alpha \beta}}{R}  ,
\label{a1-160}\\
	- \frac{1}{2} g_{\mu \nu} \left\langle
	\widehat{\delta R} \widehat{\delta g^{\mu \nu}}
	\right\rangle &=& - \frac{F^2}{2} \frac{R_{\alpha \beta }R^{\alpha \beta}}{R} +
	\frac{F}{2} \nabla^\alpha \nabla_\alpha  F -
	\frac{F}{2} \left(
		\nabla^\alpha \nabla^\beta  F \frac{R_{\alpha \beta}}{R}
	\right).
\label{a1-170}
\end{eqnarray}
Summing \eqref{a1-140}-\eqref{a1-170}, we have
\begin{equation}
\begin{split}
	\left\langle
		\widehat{\delta^2 \mathcal L}
	\right\rangle = & -\frac{c^2}{\varkappa}\Big\{
-\frac{3}{2} F^2 \frac{R_{\alpha \beta }R^{\alpha \beta}}{R} +
	F \frac{R^{\alpha \beta}}{R} \left[
		\frac{1}{2} \nabla_\alpha \nabla_\rho \left(
			F \frac{R^\rho_\beta}{R}
		\right) -
		\frac{1}{4} \nabla^\rho \nabla_\rho \left(
			F \frac{R_{\alpha \beta}}{R}
		\right) - \frac{1}{4} \nabla_\alpha \nabla_\beta F
	\right] -
\\	
	&
	\frac{F}{2} \nabla^\alpha \nabla^\beta \left(
		F \frac{R_{\alpha \beta}}{R}
	\right) + \frac{1}{2} R_\alpha^\sigma R_{\sigma \beta} R^{\alpha \beta} \frac{F^2}{R} -
	\frac{1}{2} R_{\alpha \rho \beta \sigma} R^{\rho \sigma} R^{\alpha \beta} \frac{F^2}{R^2} +
	\frac{F}{2} \nabla^\rho \nabla_\rho F +
	\frac{F^2 R^2}{4}\Big\}.
\end{split}
\label{a1-180}
\end{equation}

\section{Nonperturbative vacuum: discussion and definitions}
\label{NPvacuum}

In perturbative quantum field theories a vacuum is defined by using an annihilation operator $\hat a$ as
\begin{equation}
  \left. \left.
    \hat a \right| 0
  \right\rangle = 0.
\label{6d-10}
\end{equation}
This definition explicitly uses the notion of quantum and consequently cannot be used for the definition of a nonperturbative vacuum.
Physically, the difference between perturbative and nonperturbative vacua is the following:
the perturbative vacuum is a sea of virtual quanta that appear and annihilate everywhere and always;
the nonperturbative vacuum is similar to a stormy sea
%(nonperturbative vacuum)
with random waves (fluctuating fields) on it.

We give the following definition of a nonperturbative vacuum for gravity (here we work with tetrads):
\begin{enumerate}
  \item The expectation value of tetrad operators
  $\hat e^a_{\phantom a \mu}$ at any point $x^\mu$ should satisfy the following relation:
  \begin{equation}
    \left \langle 0 \left |
      \hat e_{a \mu} (x^\rho) \hat e^a_{\phantom a \nu} (x^\rho)
    \right| 0
    \right\rangle = \eta_{\mu \nu},
  \label{6d-60}
  \end{equation}
  where $\eta_{\mu \nu}$ is the Minkowski metric.
  \item For some combinations of indices $a,b, \mu, \nu$ the 2-point Green's function of tetrad operators $\hat e^a_{\phantom a \mu}$ is non-zero,
  \begin{equation}
    G^{a b}_{2; \rho, \sigma} \left( x^\rho, x^\sigma \right) =
    \left \langle 0 \left |
      \hat e^a_{\phantom a \mu} (x^\rho)
      \hat e^b_{\phantom b \nu} (x^\sigma)  \right| 0
    \right\rangle \neq 0 .
  \label{6d-70}
  \end{equation}
  \item The dispersion of the metric at any point $x^\mu$ is non-zero,
  \begin{equation}
    \left \langle 0 \left | \Big(
        \hat e_{a \mu} (x^\rho) \hat e^a_{\phantom a \nu} (x^\rho) -
        \eta_{\mu \nu}
       \Big)^2
      \right| 0
    \right\rangle \neq 0.
  \label{6d-80}
  \end{equation}
\end{enumerate}
\textcolor{blue}{\emph{Thus, the definition of a  nonperturbative vacuum for gravity
is obtained by combining the set of equations  for all Green's functions \eqref{6d-2-20}-\eqref{6d-2-70} with the constraints \eqref{6d-60}-\eqref{6d-80}.}
}

In order to obtain the set of equations for all Green's functions, we use the operator Einstein equations
\begin{equation}
	\hat R_{\mu \nu} - \frac{1}{2} \hat g_{\mu \nu} \hat R =
	\varkappa \hat T_{\mu \nu} .
\label{6d-2-10}
\end{equation}
The equations for all Green's functions, which follow from the operator equation, are
\begin{eqnarray}
  \left\langle 0 \left|
    \left[
      \hat R_{\mu \nu} - \frac{1}{2} \hat g_{\mu \nu} \hat R
  \right]_{x=x^\nu}
  \right| 0 \right\rangle &=& \varkappa \left\langle 0 \left|
    \hat T_{\mu \nu}
  \right| 0 \right\rangle,
\label{6d-2-20}\\
  \left\langle 0 \left|
  \hat e^{a_1}_{\phantom a \alpha_1} \left( x_1^\mu \right) \cdot
  \left[
    \hat R_{\mu \nu} - \frac{1}{2} \hat g_{\mu \nu} \hat R
  \right]_{x=x^\nu}
  \right| 0 \right\rangle &=& \varkappa \left\langle 0 \left|
    \hat e^{a_1}_{\phantom a \alpha_1} \left( x_1^\mu \right) \cdot
    \hat T_{\mu \nu}
  \right| 0 \right\rangle,
\label{6d-2-30}\\
  \left\langle 0 \left|
  \hat e^{a_1}_{\phantom a \alpha_1} \left( x_1^\mu \right) \cdot
  \hat e^{a_2}_{\phantom a \alpha_2} \left( x_2^\mu \right) \cdot
  \left[
    \hat R_{\mu \nu} - \frac{1}{2} \hat g_{\mu \nu} \hat R
  \right]_{x=x^\nu}
  \right| 0 \right\rangle &=& \varkappa \left\langle 0 \left|
    \hat e^{a_1}_{\phantom a \alpha_1} \left( x_1^\mu \right) \cdot
  \hat e^{a_2}_{\phantom a \alpha_2} \left( x_2^\mu \right) \cdot
    \hat T_{\mu \nu}
  \right| 0 \right\rangle,
\label{6d-2-40}\\
  \cdots &=& \cdots ,
\label{6d-2-50}\\
  \left\langle 0 \left|
  \hat e^{a_1}_{\phantom a \alpha_1} \left( x_1^\mu \right) \cdots
  \hat e^{a_n}_{\phantom a \alpha_n} \left( x_n^\mu \right) \cdot
  \left[
    \hat R_{\mu \nu} - \frac{1}{2} \hat g_{\mu \nu} \hat R
  \right]_{x=x^\nu}
  \right| 0 \right\rangle &=& \varkappa \left\langle 0 \left|
    \hat e^{a_1}_{\phantom a \alpha_1} \left( x_1^\mu \right) \cdots
  \hat e^{a_n}_{\phantom a \alpha_n} \left( x_n^\mu \right) \cdot
    \hat T_{\mu \nu}
  \right| 0 \right\rangle,
\label{6d-2-60}\\
  \cdots &=& \cdots
\label{6d-2-70}
\end{eqnarray}
with the constraints \eqref{6d-60}-\eqref{6d-80}.
%The set of
Equations \eqref{6d-2-20}-\eqref{6d-2-70}
are partial differential equations for all Green's functions. The solution of this set of equations  with the
constraints \eqref{6d-60}-\eqref{6d-80} gives us all Green's functions describing a vacuum state in quantum gravity.
The knowledge of all Green's functions is identical to knowing the properties of the operators $\hat e^{a}_{\phantom a \mu}$ and the vacuum quantum state~
$
\left. \left. \right| 0 \right\rangle
$.

In the same way, we can define the quantum state
$
\left. \left. \right| Q \right\rangle
$: \textcolor{blue}{\emph{The quantum state
$
\left. \left. \right| Q \right\rangle
$ and the properties of the operators $\hat e^a_{\phantom a \mu}$ are defined through all Green's functions satisfying the set of equations
}}
\begin{eqnarray}
  \left\langle Q \left|
    \left[
      \hat R_{\mu \nu} - \frac{1}{2} \hat g_{\mu \nu} \hat R
  \right]_{x=x^\nu}
  \right| Q \right\rangle &=& \varkappa \left\langle Q \left|
    \hat T_{\mu \nu}
  \right| Q \right\rangle,
\label{6d-2-80}\\
  \left\langle Q \left|
  \hat e^{a_1}_{\phantom a \alpha_1} \left( x_1^\mu \right) \cdot
  \left[
    \hat R_{\mu \nu} - \frac{1}{2} \hat g_{\mu \nu} \hat R
  \right]_{x=x^\nu}
  \right| Q \right\rangle &=& \varkappa \left\langle Q \left|
    \hat e^{a_1}_{\phantom a \alpha_1} \left( x_1^\mu \right) \cdot
    \hat T_{\mu \nu}
  \right| Q \right\rangle,
\label{6d-2-90}\\
  \left\langle Q \left|
  \hat e^{a_1}_{\phantom a \alpha_1} \left( x_1^\mu \right) \cdot
  \hat e^{a_2}_{\phantom a \alpha_2} \left( x_2^\mu \right) \cdot
  \left[
    \hat R_{\mu \nu} - \frac{1}{2} \hat g_{\mu \nu} \hat R
  \right]_{x=x^\nu}
  \right| Q \right\rangle &=& \varkappa \left\langle Q \left|
    \hat e^{a_1}_{\phantom a \alpha_1} \left( x_1^\mu \right) \cdot
  \hat e^{a_2}_{\phantom a \alpha_2} \left( x_2^\mu \right) \cdot
    \hat T_{\mu \nu}
  \right| Q \right\rangle,
\label{6d-2-100}\\
  \cdots &=& \cdots ,
\label{6d-2-110}\\
  \left\langle Q \left|
  \hat e^{a_1}_{\phantom a \alpha_1} \left( x_1^\mu \right) \cdots
  \hat e^{a_n}_{\phantom a \alpha_n} \left( x_n^\mu \right) \cdot
  \left[
    \hat R_{\mu \nu} - \frac{1}{2} \hat g_{\mu \nu} \hat R
  \right]_{x=x^\nu}
  \right| Q \right\rangle &=& \varkappa \left\langle Q \left|
    \hat e^{a_1}_{\phantom a \alpha_1} \left( x_1^\mu \right) \cdots
  \hat e^{a_n}_{\phantom a \alpha_n} \left( x_n^\mu \right) \cdot
    \hat T_{\mu \nu}
  \right| Q \right\rangle,
\label{6d-2-120}\\
  \cdots &=& \cdots
\label{6d-2-130}
\end{eqnarray}
Let us emphasize that for the definition of the quantum state
$
\left. \left. \right| Q \right\rangle
$ we do not use the constraints \eqref{6d-60}-\eqref{6d-80}.

Finally, let us address the procedure of quantum averaging. In quantum mechanics, this procedure is carried out by the integration:
\begin{equation}
	\left\langle \hat L \right\rangle = \int \psi^* \hat L \psi dV
\label{6d-2-140}.
\end{equation}
For the nonperturbative quantization, this procedure is not obvious.
%In our case
To define a quantum average for an operator, or some product of operators,
we first have to solve the set of equations  \eqref{6d-2-80}-\eqref{6d-2-130}. Then,  among the obtained Green's functions, we find the required quantum average.

\end{document}